\documentclass[fleqn,usenatbib]{mnras}
 
\usepackage{graphicx} 
\usepackage{txfonts} 
\usepackage{natbib}               
\title[]{{\it Hubble} Space Telescope survey of Magellanic Cloud star clusters. UV-dim stars in young clusters}

\author[ A.\,P.\,Milone et al.] 
       {A.\,P.\,Milone$^{1,2}$, 
       G.\,Cordoni$^{2}$, 
       A.\,F.\,Marino$^{2,3}$, 
       F.\,Muratore$^{1}$, 
       F.\,D'Antona$^{4}$,
       M.\,Di Criscienzo$^{4}$,  \newauthor
        E.\,Dondoglio$^{1}$, 
       E.\,P.\,Lagioia$^{1}$,
       M.\,V.\,Legnardi$^{1}$, A.\,Mohandasan$^{1}$, T.\,Ziliotto$^{1}$,
       F.\,Dell'Agli$^{1}$,
          \newauthor M.\,Tailo$^{5}$, 
        P.\,Ventura$^{4}$ 
        \\
$^{1}$Dipartimento di Fisica e Astronomia ``Galileo Galilei'', Univ. di Padova, Vicolo dell'Osservatorio 3, Padova, IT-35122\\
$^{2}$Istituto Nazionale di Astrofisica - Osservatorio Astronomico di Padova, Vicolo dell'Osservatorio 5, Padova, IT-35122\\
$^{3}$  Istituto Nazionale di Astrofisica - Osservatorio Astrofisico di Arcetri, Largo Enrico Fermi, 5, Firenze, IT-50125 \\
$^{4}$ INAF - Osservatorio Astronomico di Roma, Via Frascati 33, I-00040, Monte Porzio Catone, Roma, Italy \\
$^{5}$ Dipartimento di Fisica e Astronomia Augusto Righi, Università degli Studi di Bologna, Via Gobetti 93/2, 40129, Bologna, Italy\\
} 
\begin{document}

\maketitle 
\label{firstpage}
\begin{abstract}
Young and intermediate-age star clusters of both Magellanic Clouds exhibit complex color-magnitude diagrams. In addition to the extended main-sequence turn-offs (eMSTOs), commonly observed in star clusters younger than $\sim$2 Gyr, the clusters younger than $\sim$800 Myr exhibit split main sequences (MSs). These comprise a blue MS, composed of stars with low-rotation rates, and a red MS, which hosts fast-rotating stars.
 While it is widely accepted that stellar populations with different rotation rates are responsible for the eMSTOs and split MSs, their formation and evolution are still debated.
 A recent investigation of the $\sim$1.7 Gyr old cluster NGC\,1783 detected a group of eMSTO stars extremely dim in UV bands.
Here, we use multi-band {\it Hubble Space Telescope} photometry to investigate five star clusters younger than $\sim$200\,Myr, including   NGC\,1805, NGC\,1818, NGC\,1850, and NGC\,2164 in the Large Magellanic Cloud, and the Small-Magellanic Cloud cluster NGC\,330. 
We discover a group of bright MS stars in each cluster that are significantly dim in the F225W and F275W bands, similar to what is observed in NGC\,1783. 
Our result suggests that UV-dim stars are common in young clusters. The evidence that most of them populate the blue MS indicates that they are slow rotators. As a byproduct, we show that the star clusters NGC\,1850 and BHRT\,5b exhibit different proper motions, thus corroborating the evidence that they are not gravitationally bound.
\end{abstract}


\section{Introduction}\label{sec:intro}
In recent years, the split main sequence (MS) has been discovered as a common feature of young-cluster color-magnitude diagrams  \citep[CMDs,][]{Milone2015, milone2023a}. High-precision photometry from {\it Hubble Space Telescope} ({\it HST}) images reveals that all Magellanic Cloud clusters younger than $\sim$800 Myr exhibit a blue MS, which comprises about one-third of MS stars, and a more-populous red MS\citep[e.g.][]{Milone2018}. 

The blue MS consists of fast rotators, whereas blue-MS stars exhibit small rotational velocities, as demonstrated by high-resolution spectra of MS stars \citep{Marino2018a, Marino2018b}. 
These findings corroborate the conclusion based on  isochrone fitting of the CMD that stellar rotation is the main responsible for the split MS \citep{Dantona2015a}. 

Similarly, the extended main-sequence turn-off (eMSTO), a distinctive feature of star clusters younger than $\sim 2$Gyr \citep{Mackey2007, Milone2009, goudfrooij2011},
 is populated by stars with different rotation rates as early suggested by theoretical studies \citep{Bastian2009, Dantona2015a, georgy2019a} and demonstrated by works based on spectroscopy \citep{Dupree2017, Marino2018a, Marino2018b}.
 However, rotation alone does not seem enough to fully account for the observed eMSTO.
 Hence, it is possible that age variation provides a contribution 
 (e.g.\citet[][]{goudfrooij2017a}, but  see \citet{cordoni2022a}).
Intriguingly, split or broadened MSs and eMSTOs have been observed in Galactic open clusters with the same ages \citep{Marino2018a, Cordoni2018a, li2019a}.

The origin of multiple stellar populations with different rotation rates is still under debate.  \citet{Dantona2015a, DAntona2017a} suggested that all stars are born as fast rotators and that some of them are braked during their MS lifetime due to interaction in binary systems. 
 As an alternative, the different rotation velocities may be imprinted in the early time of cluster formation depending on the lifetime of proto-stellar discs in pre-MS stars. Specifically, present-day fast rotators are the progeny of pre-MS stars with short-lived proto-stellar discs, whereas long removal times for proto-stellar discs would result in slow rotators \citep{Tailo2016a, Bastian2020a}.
 \citet{wang2022a} suggested that the blue-MS stars are slow rotators originating from stellar mergers. Their scenario predicts that red-MS stars gain their mass by disk accretion, which leads to rapid rotation, whereas binary merger is responsible for the formation of the blue-MS.
 
Recently, a new CMD feature has been discovered along the eMSTO of NGC\,1783. In the first paper of this series, \citet{milone2023a} have detected  a group of stars located on the red side of the eMSTO in CMDs composed of ultraviolet filters \citep[see also][]{milone2022b}. These UV-dim stars, which comprise $\sim$7\% of the total number of eMSTO stars of NGC\,1783, define a narrow sequence and have 
 colors similar to the remaining eMSTO stars, when observed in optical CMDs.

The observed colors and magnitudes of UV-dim stars can be reproduced 
 by assuming the presence of an edge-on absorption dust ring, 
associated with grain condensation at the periphery of the excretion disc expelled by fast-spinning stars \citep{dantona2023a}.

In this scenario, the entire eMSTO could be composed of dusty stars, which would significantly affect previous conclusions on the role of age and rotation on the eMSTO \citep[see][for details]{dantona2023a}.

\citet{martocchia2023a} have recently suggested that UV-dim stars are likely rapidly rotating stars. They associated the shell Be-stars of the $\sim$100 Myr old Large Magellanic Cloud (LMC) cluster NGC\,1850 that are seen nearly
equator-on  with the UV-dim stars of this cluster.

In this work, we use multi-band {\it HST} photometry to search for UV-dim stars in five Magellanic-Cloud star clusters with ages between $\sim$10 and 200 Myr, namely NGC\,330, NGC\,1805, NGC\,1818, NGC\,1850, and NGC\,2164 \footnote{ According to \citet{Milone2018}, who using the isochrones from the Geneva database \citep{mowlavi2012a, ekstrom2013a, Ekstrom2012}, the ages of the studied clusters span the range between $\sim$30 to 100 Myr. The isochrones from the Padova database \citep{marigo2017a} provide ages between $\sim 10$ Myr and 200 Myr \citep{milone2023a}. The latter value may represent an upper limit to the age of NGC\,2164 and is derived by assuming that the color broadening of eMSTO stars is entirely due to age variations. For simplicity, we indicate the value of 200 Myr as an upper limit for the age of the studied clusters.}. 
 These clusters, which are widely investigated in the context of multiple populations, exhibit split MSs, eMSTOs, and conspicuous populations of Be stars \citep[]{li2017, correnti2017a, bastian2017a, Milone2018, milone2023a, Marino2018a, hastings2021a, cordoni2022a}. Our targets are the only young clusters with available photometry in the F225W or F275W bands.

The paper is organized as follows. In Section\,\ref{sec:data} we describe the data and the data reduction. Section\,\ref{sec:UVn2146} is dedicated to the search for UV-dim stars in NGC\,2164, which we use as a test case, whereas the remaining clusters are analyzed in Section\,\ref{sec:clusters}. Finally, the summary and the conclusions are provided in Section\,\ref{sec:summary}.
As a byproduct of our investigation, the Appendix presents the proper motions of the double cluster composed of NGC\,1850 and HRBT\,5b and of the LMC stars in their field of view.

\section{Data and data analysis}\label{sec:data}
To study the young clusters, we used the  catalogs by \citet{milone2023a}, which include astrometry and photometry of 113 Magellanic Cloud clusters derived from {\it HST} images. 
We used photometry in the F336W and F814W bands of the Ultraviolet and Visual Channel of the Wide Field Camera 3 (UVIS/WFC3) of all clusters. In addition, we used F225W UVIS/WFC3 photometry of NGC\,330, NGC\,1805, NGC\,1818, and NGC\,2164, and F275W UVIS/WFC3 photometry of NGC\,1850.
In addition, we reduced and analyzed the UVIS/WFC3 images of NGC\,1850 collected on September 21$^{st}$, 2021 as part of GO\,16748 (P.\,I.\, F.\,Niederhofer). These data, which consist of two 40s images and five 350s images observed through the F814W filter, are used to measure the proper motions of stars in the field of view of NGC\,1850.

Stellar fluxes and positions are measured by \citet{milone2023a} with the KS2 computer program, which was developed by Jay Anderson as  the evolution of the $kitchen\_sync$ \citep{Anderson2008}, originally written to reduce images collected with the Wide Field Channel of the Advanced Camera for Surveys onboard {\it HST}. In a nutshell, KS2 uses three methods to measure the stars, which derive the optimal astrometry and photometry for stars with different luminosities. Since we are interested in bright stars, we used the results obtained from method I. It measures all stellar sources that generate a distinct peak within a 5$\times$5 pixel region after neighbor stars subtraction. To do this, KS2 calculates the flux and position of each star in each image, separately, by using the point-spread function (PSF) model associated with its position and subtracting the sky level computed from the annulus between four and eight pixels from the stellar center. Finally, the results from all images are averaged together to obtain the best determinations of stellar magnitudes and positions.
The KS2 computer program provides various diagnostics of the photometric quality. Since we are interested in high-precision photometry, we included in the analysis only isolated stars that are well-fitted by the PSF model \citep[see section 2.4 from][for details]{milone2023a}.
For NGC\,1850, which is the only analyzed cluster that is  significantly affected by differential reddening, we used the magnitudes corrected for differential reddening provided by \citet{milone2023a}.
To derive the proper motions of stars in the field of view of NGC\,1850, we followed the procedure by \citet[][see their section 5]{milone2023a}. In a nutshell, we first derived proper motions relative to NGC\,1850 by  comparing the positions of stars at different epochs. Specifically, in addition to the GO-16748 images, we used UVIS/WFC3 images collected in F814W on October 22, 2015, as part of GO-14174, F438W images observed on December 19, 2015 (GO-14069), and F467M images taken on October 20, 2009 (GO-11925). Stellar positions are corrected for geometric distortion by using the solution provided by \citet{bellini2009a} and \citet{bellini2011a}. The proper motions are transformed from relative to absolute by using the stars for which both relative proper motions and absolute proper motions from the Gaia data release 3 \citep{gaia2022a} are available \citep[see][for details on the dataset and on the methods to estimate the proper motions]{milone2023a}.

\begin{figure} 
\centering
\includegraphics[width=9cm,trim={0.7cm 5.1cm 0.2cm 4.5cm},clip]{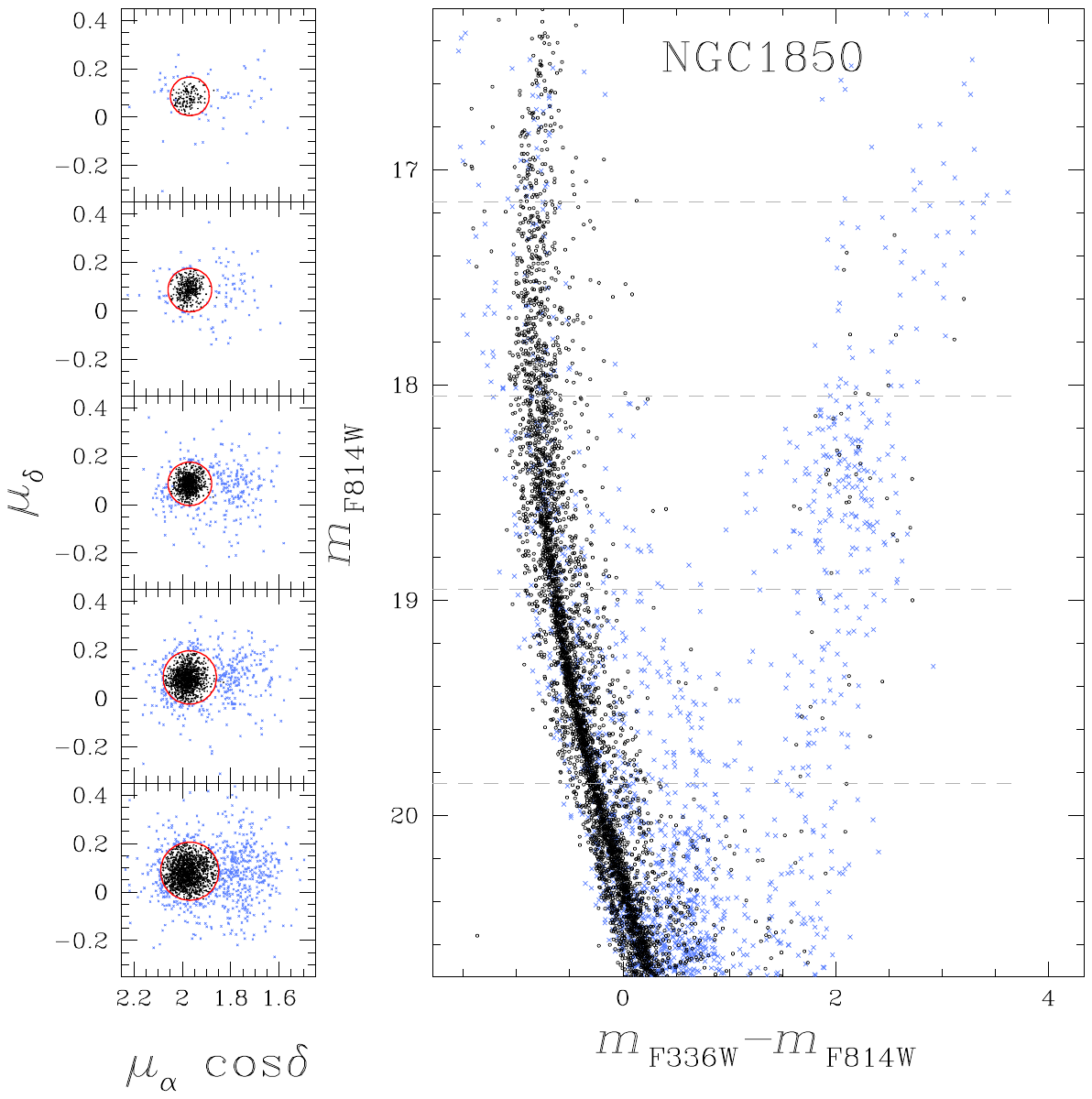}
 \caption{Proper-motion diagram of stars in the field of view of NGC\,1850 in five magnitude intervals (left). The right panel shows the $m_{\rm F814W}$ vs.\,$m_{\rm F336W}-m_{\rm F814W}$ CMD for stars with available proper motions. The probable cluster members (i.e.\, stars within the red circles of the left panels) are colored black, whereas the azure symbols represent the remaining stars. }
 \label{fig:ngc1850pm} 
\end{figure}

The results are illustrated in Figure\,\ref{fig:ngc1850pm}, where we plot the proper-motion diagrams for stars in the field of view of NGC\,1850 in five F814W magnitude intervals.
 The proper motions allowed us to separate the majority of cluster members (black dots) from  probable field stars (azure crosses). The red circles that enclose the bulk of cluster members, are centered on the average motion of NGC\,1850 (($\mu_{\alpha} cos \delta$, $\mu_{\delta}$)=(1.973$\pm$0.026,0.086$\pm$0.028) mas/yr) and have radii equal to 2.5$\sigma$, where $\sigma$ is the average between the $\sigma-$clipped dispersion values of $\mu_{\alpha} {\rm cos} \delta$ and $\mu_{\delta}$.  
 The right panel of Figure\,\ref{fig:ngc1850pm} shows the  $m_{\rm F814W}$ versus\,$m_{\rm F336W}-m_{\rm F814W}$ CMD of the selected cluster members and field stars.

To investigate each cluster, we analyze the stars within two circular regions defined by \citet{Milone2018}  within the same UVIS/WFC3 field of view. They include the cluster field, which is centered on the cluster and is mostly populated by cluster members, and the reference field, which has the same area as the cluster field but is mostly populated by field stars. The latter,  which is located in the outskirts of the UVIS/WFC3 field, is used to
statistically estimate the contamination of field stars in the cluster field  \citep[see][for details]{Milone2018}.

\subsection{Artificial stars}
To estimate the photometric errors and generate the simulated CMD, we performed artificial star (AS) tests for each cluster. To do this, we followed the recipe by \citet{Anderson2008} and derived a list of 99,999 artificial stars with similar radial distributions and luminosity functions as the observed stars. 
  The ASs have instrumental magnitudes\footnote{The instrumental magnitudes are defined as $-2.5 \log_{10}$ of the detected photo-electrons.} between the saturation limit of $\sim -$13.7 mag and $-$5.0 mag in the F814W UVIS/WFC3 filter. The magnitudes in the other filters are derived from the fiducial lines of the red and blue MS.
To derive the magnitudes and positions of the ASs we used the KS2 program and the same procedure that \citet{milone2023a} adopted for real stars. 
 We included in our investigation only the relatively isolated ASs that are well-fitted by the PSF and that we selected by using the same criteria adopted for the real stars.

\section{UV-dim stars in NGC\,2164}
\label{sec:UVn2146}
To search for UV-dim stars in young clusters, we used the $\sim 200$\,Myr-old LMC cluster NGC\,2164 as a test case.
We first exploited the intermediate-age ($\sim$1.7 Gyr old) LMC cluster NGC\,1783, where UV-dim stars have been discovered by \cite{milone2022b} to constrain the properties of these stars in 
photometric diagrams constructed with optical and UV filters. 
Then, we extended the analysis to NGC\,2164 and the other young clusters studied in this paper.

\begin{figure*} 
\centering
\includegraphics[width=11cm,trim={0.0cm 5.1cm 0.0cm 4.5cm},clip]{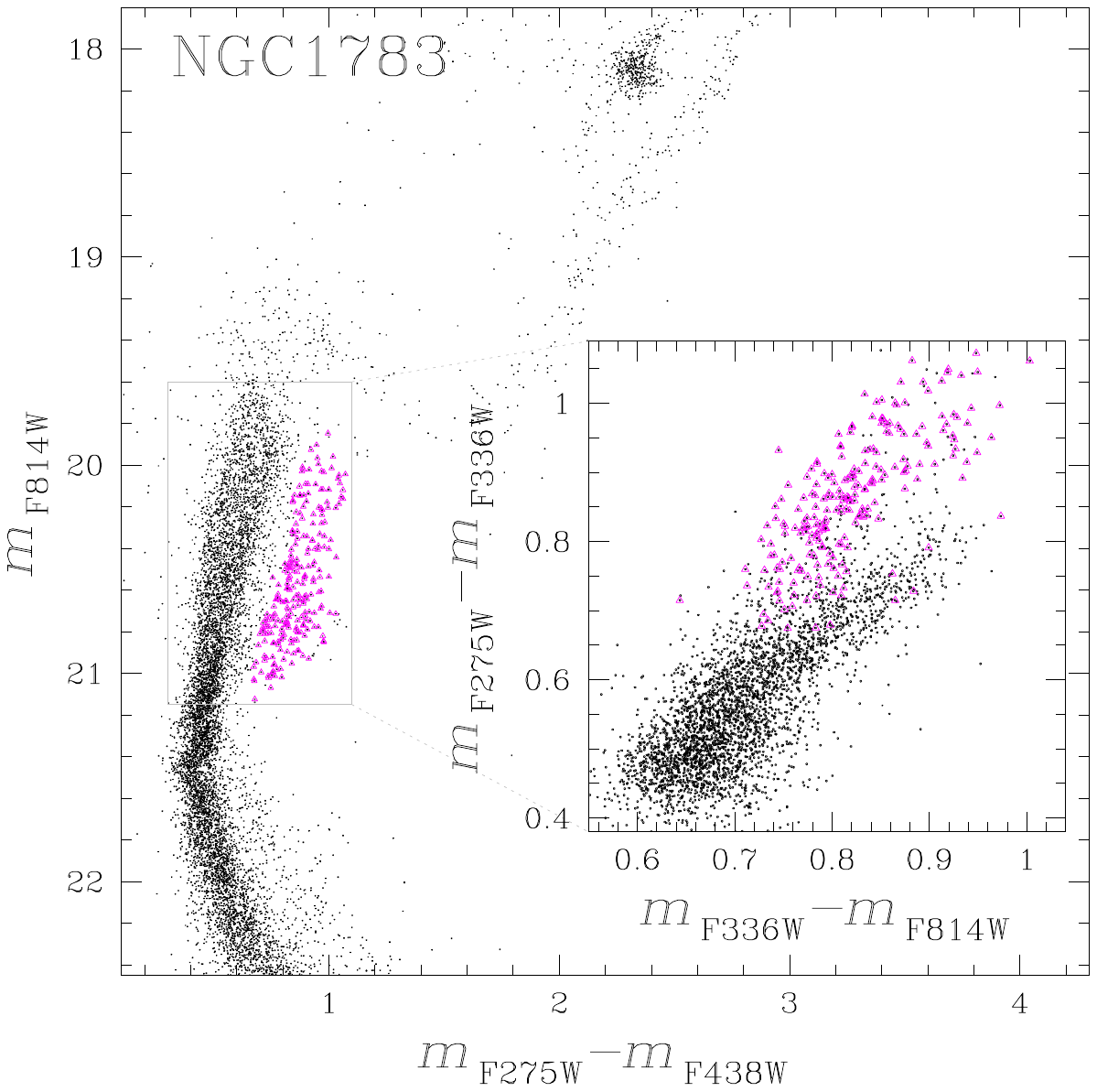}
 \caption{$m_{\rm F814W}$ vs.\,$m_{\rm F275W}-m_{\rm F438W}$ CMD of the $\sim$1.7-Gyr old LMC cluster NGC\,1783. The inset shows the $m_{\rm F275W}-m_{\rm F336W}$ vs.\,$m_{\rm F336W}-m_{\rm F814W}$ two-color diagram for eMSTO stars. The UV-dim stars selected by \citet{milone2023a} are marked with magenta triangles.}
 \label{fig:ngc1783uvdim} 
\end{figure*}

As illustrated in Figure\,\ref{fig:ngc1783uvdim}, the $m_{\rm F814W}$ vs.\,$m_{\rm F275W}-m_{\rm F438W}$ CMD of NGC\,1783 exhibits a cloud of stars with redder colors than the bulk of eMSTO stars. The most-evident UV-dim stars, selected by \citet{milone2023a}, are marked with magenta triangles. We plotted in the inset of Figure\,\ref{fig:ngc1783uvdim} the $m_{\rm F275W}-m_{\rm F336W}$ vs.\,$m_{\rm F336W}-m_{\rm F814W}$ two-color diagram for the stars enclosed in the gray box. Clearly, UV-dim stars define a different sequence with respect to the remaining eMSTO stars, and exhibit redder $m_{\rm F275W}-m_{\rm F336W}$ colors than the bulk of eMSTO stars with similar $m_{\rm F336W}-m_{\rm F814W}$ values.
 Hence, this two-color diagram is an efficient tool to identify the UV-dim stars.

\begin{figure*} 
\centering
\includegraphics[width=12.00cm,trim={0.0cm 5.5cm 0.0cm 10.75cm},clip]{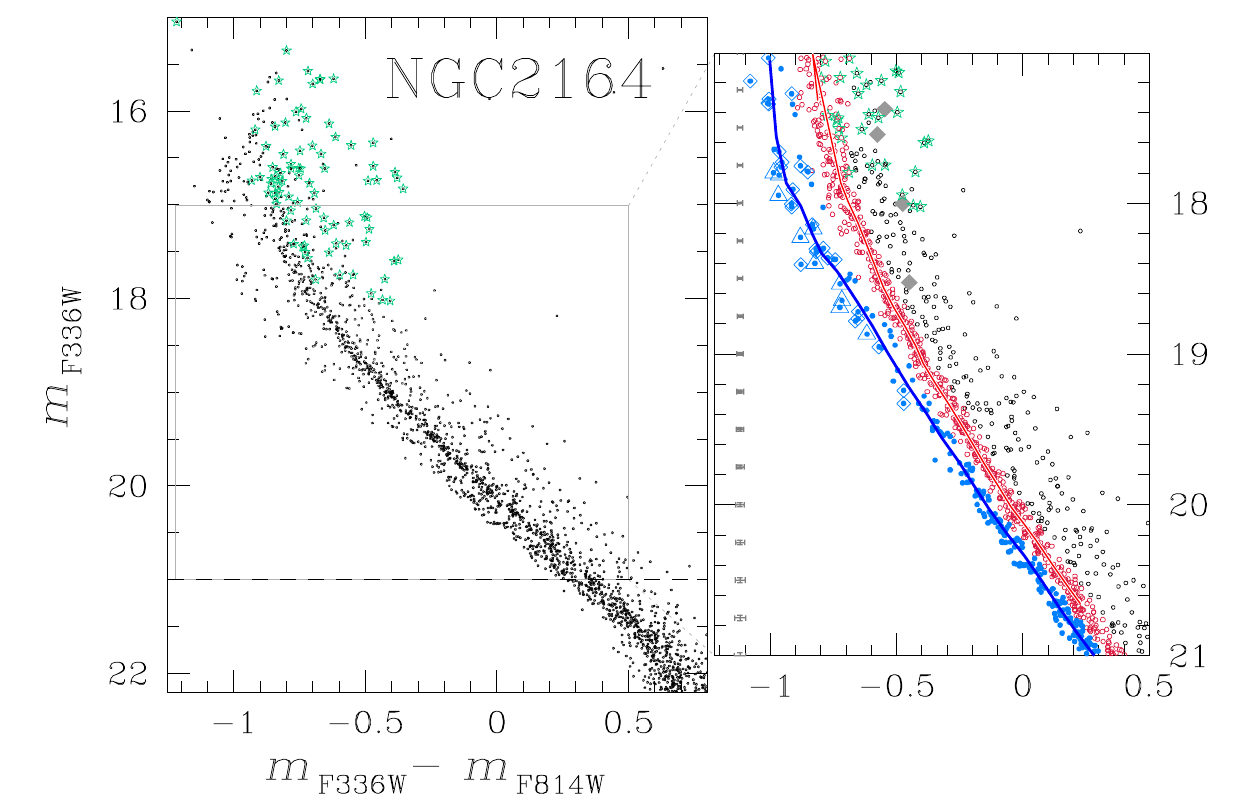}
\includegraphics[width=12.00cm,trim={0.0cm 5.5cm 0.0cm 10.75cm},clip]{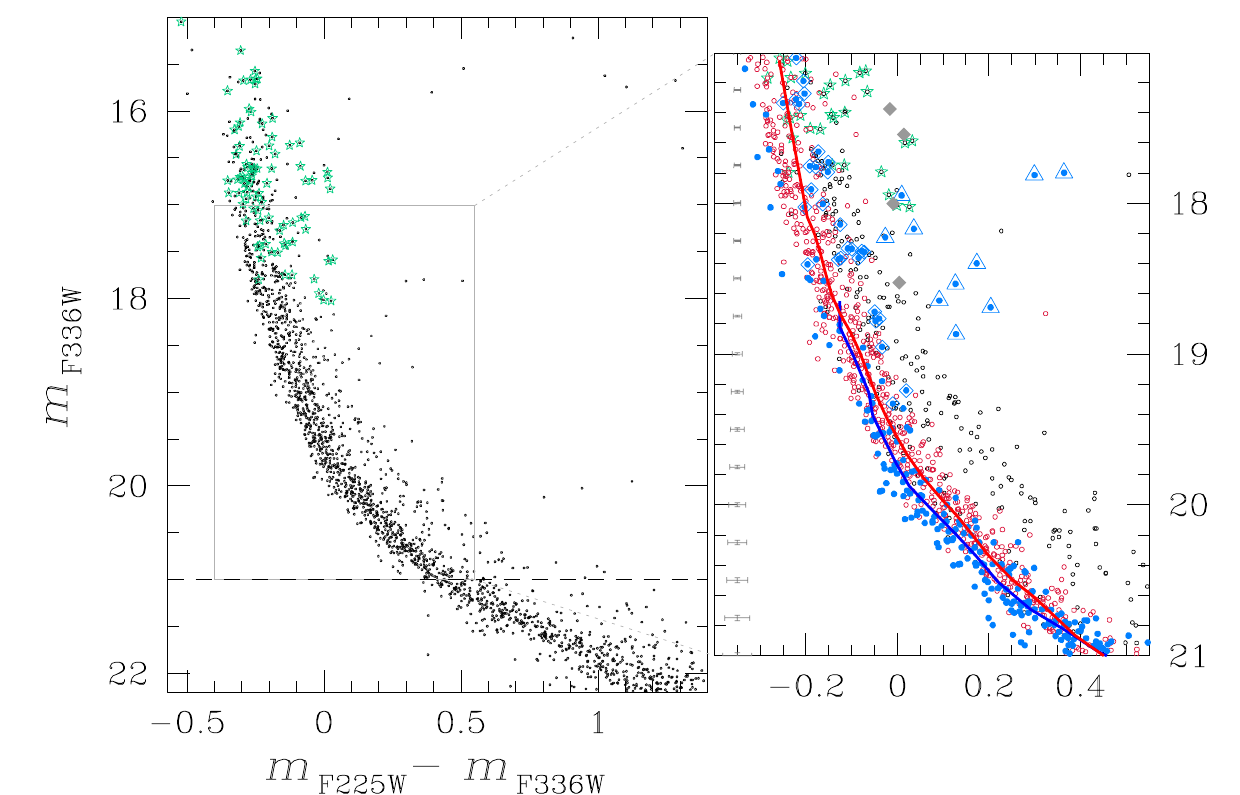}
\includegraphics[width=11.5cm,trim={1.0cm 5.5cm 1.5cm 14.65cm},clip]{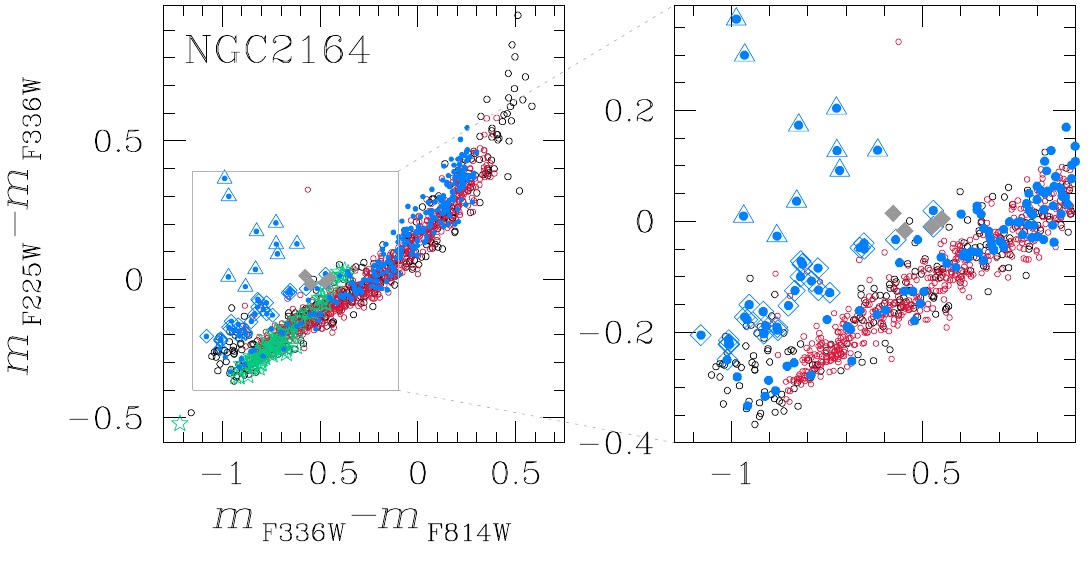}
 \includegraphics[width=4.48cm,trim={12.3cm 5.5cm 1.5cm 14.65cm},clip]{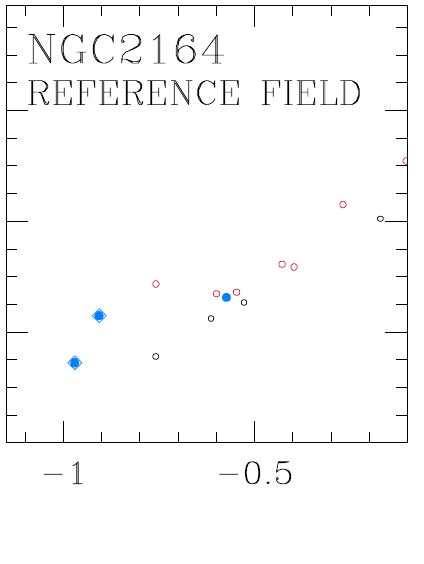}
 \caption{$m_{\rm F336W}$ vs.\,$m_{\rm F336W}-m_{\rm F814W}$ (top) and $m_{\rm F336W}$ vs.\,$m_{\rm F225W}-m_{\rm F336W}$ (middle) CMDs of stars in the cluster field of NGC\,2164. The right panels are zooms around the upper MS. The fiducial lines of the red and blue MSs are represented with red and blue lines, respectively. The bottom-left and bottom-middle panels show the $m_{\rm F225W}-m_{\rm F336W}$ vs.\,$m_{\rm F336W}-m_{\rm F814W}$ two-color diagrams for NGC\,2164.
  Red and blue symbols indicate red and blue-MS stars, respectively,  selected from the top-panel CMDs. 
  The Be stars identified by \citet{Milone2018} are represented with aqua-starred symbols, whereas the candidate UV-dim stars are marked with blue diamonds and triangles. Gray diamonds represent probable UV-dim stars  that, based on the position in the top-right CMD, are probable binaries. The bottom-left panel represents all the stars, while the bottom-middle panel is a zoom around the bright MS. For clearness, we excluded the Be stars from the bottom-middle panel plot.
 The $m_{\rm F225W}-m_{\rm F336W}$ vs.\,$m_{\rm F336W}-m_{\rm F814W}$ two-color diagram for stars in the reference field of NGC\,2164 is plotted in the bottom-right panel.
  }
 \label{fig:cmds} 
\end{figure*}

The top-left panel of Figure\,\ref{fig:cmds} shows the $m_{\rm F336W}$ vs.\,$m_{\rm F336W}-m_{\rm F814W}$ CMD of NGC\,2164 \citep{milone2023a}. The top-right panel provides a zoom around the CMD region where the split MS is more evident. From this diagram, we identified by eye the groups of blue and red-MS stars and colored them blue and red, respectively. The remaining stars, which are located on the red side of the MS and  are colored black, are probable binary systems composed of two MS stars. 
The blue and red lines are the fiducials of the two MSs. To derive the red fiducial, 
 we considered 0.2 mag bins defined over a grid of points separated by steps of 0.05 mag \citep{silverman1986a}. We computed the median of the colors and magnitudes of the red-MS stars in each bin and smoothed the median points with the boxcar averaging method, which replaces each point with the average of the three adjacent points. We applied the same method for deriving the blue-MS fiducial. 

The middle panels show the $m_{\rm F336W}$ vs.\,$m_{\rm F225W}-m_{\rm F336W}$ CMDs. 
As highlighted by the fiducial lines, blue- and red-MS stars exhibit similar colors. The $m_{\rm F225W}-m_{\rm F336W}$ color separation is always smaller than $\sim 0.05$ mag and the two sequences merge together around $m_{\rm F336W}=18.7$ mag, where blue-MS stars span a wide color range. 
For completeness, we used aqua-starred symbols to represent the Be stars identified by \citet{Milone2018}, based on their flux excess in the F656N band. 

To further investigate the color distribution of MS stars, we show in the bottom-left and bottom-middle panels of Figure\,\ref{fig:cmds} the $m_{\rm F225W}-m_{\rm F336W}$ vs.\,$m_{\rm F336W}-m_{\rm F814W}$ two-color diagram for the NGC\,2164 stars plotted in the top panels of Figure\,\ref{fig:cmds}. For simplicity, we only show MS and eMSTO stars brighter than $m_{\rm F336W}=21.0$, which is the magnitude level indicated by the dash-dotted lines in the top-left and middle-left panels.  

The red MS defines a narrow sequence, similar to what is observed for the blue MS with $m_{\rm F336W}-m_{\rm F814W}$  $\gtrsim -0.2$ mag. Conversely, the blue-MS stars with  $m_{\rm F336W}-m_{\rm F814W} \lesssim -0.2$ mag span a wide range of $m_{\rm F225W}-m_{\rm F336W}$ values. Specifically, we notice that about half of the blue-MS stars only  are superimposed on the red-MS. The remaining blue-MS stars (marked with open blue symbols in Figure\,\ref{fig:cmds}) exhibit larger  $m_{\rm F225W}-m_{\rm F336W}$ values than red-MS stars with the same $m_{\rm F336W}-m_{\rm F814W}$ color, in close analogy with what is observed for UV-dim stars in NGC\,1783. 
This sub-sample of blue-MS stars,  which we identify in the  $m_{\rm F225W}-m_{\rm F336W}$ vs.\,$m_{\rm F336W}-m_{\rm F814W}$ two-color diagram, includes the most-evident UV-dim stars 
 (blue triangles), which exhibit the widest color separation from the bulk of MS stars  of more than 0.20 mag. 
 The remaining UV-dim candidates with color distances larger than 0.09 mag from the fiducial line of the bulk of MS stars are marked with blue diamonds. 
 We notice that the UV-candidates include four stars that, based on the position in the top-panel CMD are binary systems composed of two MS stars (gray diamonds) \footnote{ We emphasize that F438W data are not available for NGC\,2164 and for most studied clusters, whereas photometry in the F225W (or F275W), F336W, and F814W is available for all of them. For this reason, we used the $m_{\rm F225W}-m_{\rm F336W}$ vs.\,$m_{\rm F336W}-m_{\rm F814W}$ two-color diagram to identify the candidate UV-dim stars.}.
Intriguingly, the selected Be-stars of NGC\,2164  populate a narrow sequence and span the $m_{\rm F336W}-m_{\rm F814W}$ color interval between $\sim -$0.9 and $\sim -$0.3 mag. Hence, they occupy a different region of the two-color diagram with respect to the UV-dim stars. 

A possible explanation is a consequence of the fact that the upper MS (i.e. the MS region populated by UV-dim stars) exhibits different slopes in the CMDs, and in the two-color diagrams of young clusters and intermediate-age clusters.
Specifically, the upper MS of the intermediate-age cluster NGC1783 (black stars in the inset of Figure 2) has a positive slope in the two-color diagrams, whereas the upper MS of NGC2164 and the other young clusters have negative slopes (e.g. bottom-left panels of Fig. 3 for NGC2164). 
Such slopes are consistent with those predicted by the isochrones (e.g. Figure 4).

To estimate the contamination from field stars in the two-color diagram of NGC\,2164, we analyze the same diagram for stars in the reference field. The results are illustrated in the bottom-right panel of Figure\,\ref{fig:cmds}, where the blue and red colors represent the stars that occupy the regions of the $m_{\rm F336W}$ vs.\,$m_{\rm F336W}-m_{\rm F814W}$ CMD, where we identified blue- and red-MS stars. 
 The lack of reference-field stars in the portion of the two-color diagram populated by candidate UV-dim stars indicates negligible  contamination from field stars.

\subsection{Comparison with simulated photometry}
We used the Geneva models \citep{Ekstrom2012, ekstrom2013a, mowlavi2012a} to simulate the photometric diagrams for two distinct stellar populations with the same age, in close analogy with what we did in previous work \citep[e.g.][]{Dantona2015a, milone2023a}. The simulated populations comprise 25\% of non-rotating stars, whereas the remaining stars have rotation rates corresponding to 0.9 times the critical value,  which are the rotation values that provide the best match between the Geneva models and the split MSs of NGC\,2164 \citep{Milone2018} Similarly, the adopted ages (100 Myr), and the fractions of non-rotating stars are comparable with those of NGC\,2164 \citep{Milone2018}.

 We adopted random viewing-angle distributions, the gravity-darkening model by \citet{espinosa2011a}, and the limb-darkening effect \citep{claret2000a}.
 Finally, we derived the magnitudes into the F225W, F275W, F336W, and F814W bands of UVIS/WFC3 by means of the model atmospheres by \citet{castelli2005a}.
 We adopted for the simulated stars the same photometric errors derived from  ASs with similar magnitudes and radial distances from the cluster center.

\begin{figure} 
\centering
\includegraphics[width=9cm,trim={1.5cm 5.1cm 0.5cm 12.0cm},clip]{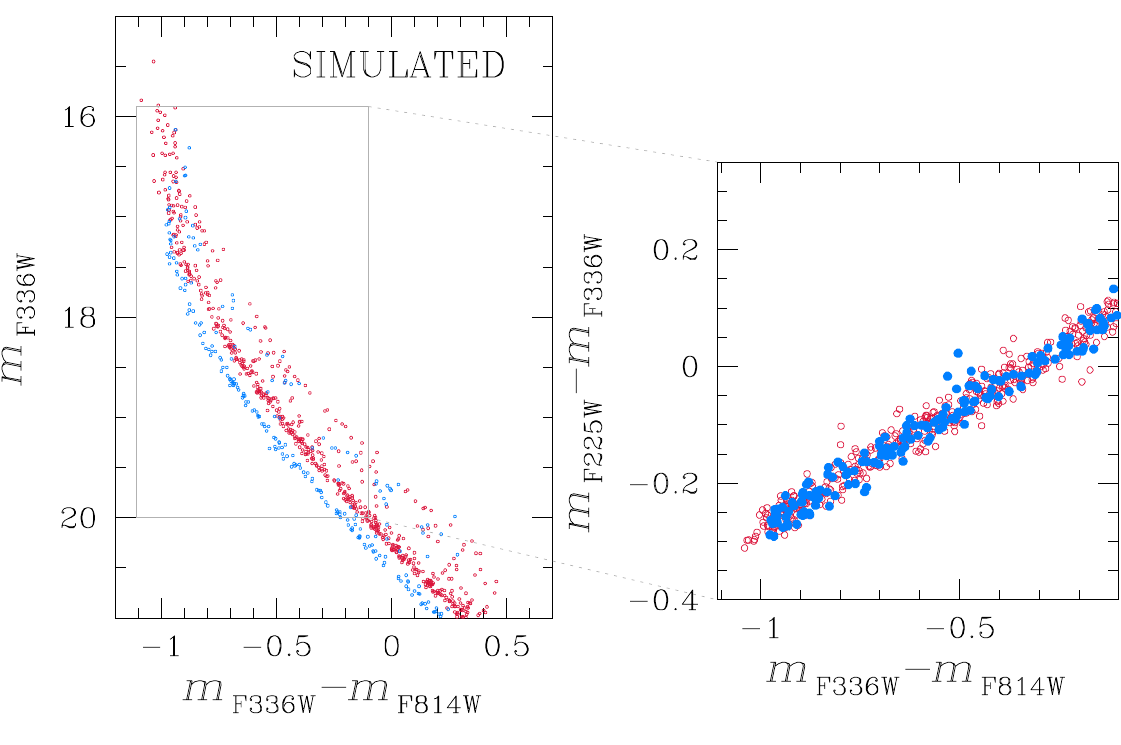}
 \caption{$m_{\rm F336W}$ vs.\,$m_{\rm F336W}-m_{\rm F814W}$ simulated CMD for a 100\,Myr old cluster (left panel). The $m_{\rm F225W}-m_{\rm F336W}$ vs.\,$m_{\rm F336W}-m_{\rm F814W}$ diagram for the MS stars shown in the CMD inset is plotted in the right panel.}
 \label{fig:simulation} 
\end{figure}

 The results are illustrated in Figure\,\ref{fig:simulation}, where we show the simulated $m_{\rm F336W}$ vs.\,$m_{\rm F336W}-m_{\rm F814W}$ CMD and the $m_{\rm F225W}-m_{\rm F336W}$ vs.\,$m_{\rm F336W}-m_{\rm F814W}$ two-color diagram for bright MS stars. The simulated stellar populations  overlap with each other in the two-color diagram.  Similar conclusions are obtained from the simulation of 20\,Myr and 200\,Myr old clusters and by assuming rotation rates equal to 0.6 times the critical value. These results indicate that the  $m_{\rm F225W}-m_{\rm F336W}$ color excess of the candidate UV-dim stars is due neither to observational errors nor to stellar rotation.

 Noticeably, a visual inspection of Figure\,\ref{fig:ngc1783uvdim} and \ref{fig:cmds} reveals that the UV-dim stars of NGC\,1783 and NGC\,2164 occupy different regions of the two-color diagrams.
The difference is possibly a consequence of the fact that the upper MS exhibits different slopes in the two-color diagrams of young clusters and intermediate-age clusters.
Specifically, the upper MS of the intermediate-age cluster NGC\,1783 (black stars in the inset of Figure\,\ref{fig:ngc1783uvdim}) has a positive slope in the two-color diagrams, whereas the upper MS of NGC\,2164 exhibits a negative slope, as predicted by the simulated diagrams of  Figure\,\ref{fig:simulation}. 
Since the UV dim stars have low fluxes in the F225W and F275W bands, compared with the remaining stars at similar evolutionary stage, they have larger F225W$-$F336W (or F275W$-$F336W) colors than the bulk of MS stars. Hence, the UV-dim stars populate the top-right part of the two-color diagram in the intermediate-age cluster NGC\,1783 but the top-left region of the diagrams of the young cluster NGC\,2164.

\section{UV-dim stars in young clusters}\label{sec:clusters}

\begin{figure*} 
\centering
\includegraphics[height=6.45cm,trim={0.0cm 5.5cm 1.78cm 14.65cm},clip]{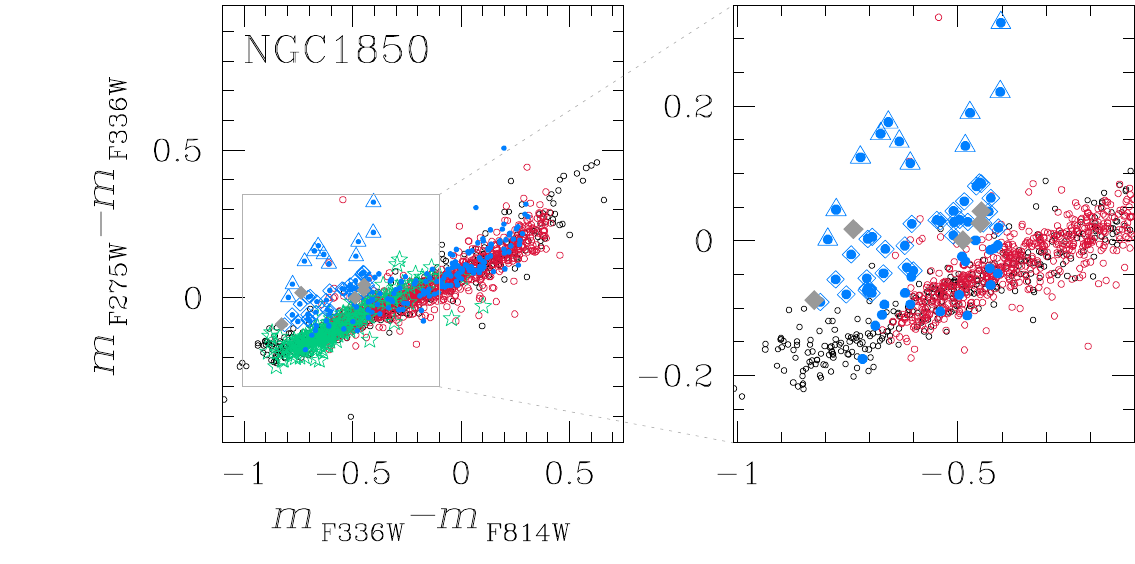}
\includegraphics[height=6.45cm,trim={12.3cm 5.5cm 1.78cm 14.65cm},clip]{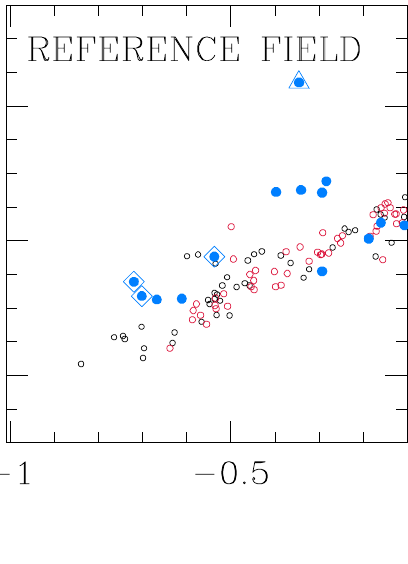}
\includegraphics[height=6.45cm,trim={0.0cm 5.5cm 1.78cm 14.65cm},clip]{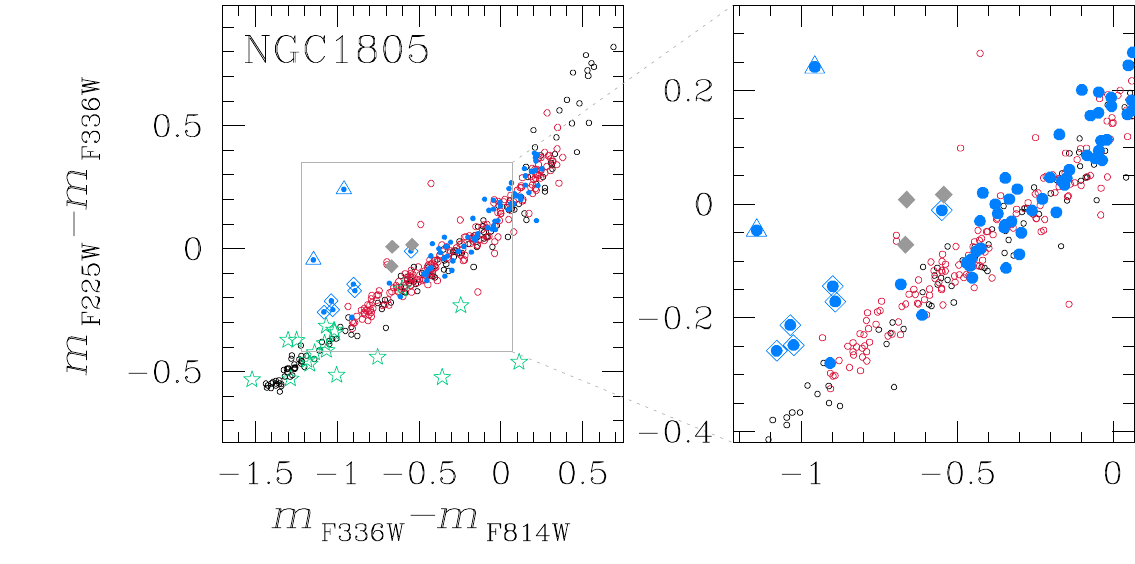}
\includegraphics[height=6.45cm,trim={12.3cm 5.5cm 1.78cm 14.65cm},clip]{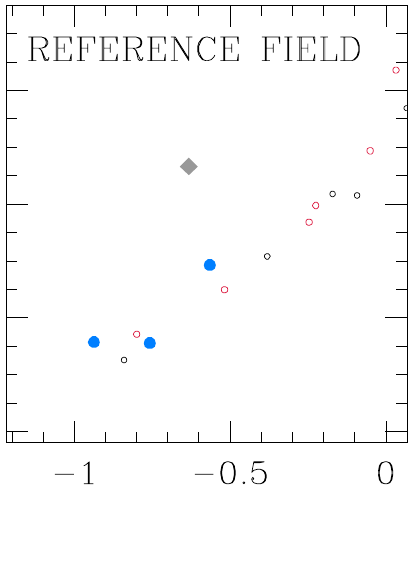}
 \includegraphics[height=6.45cm,trim={0.0cm 5.5cm 1.78cm 14.65cm},clip]{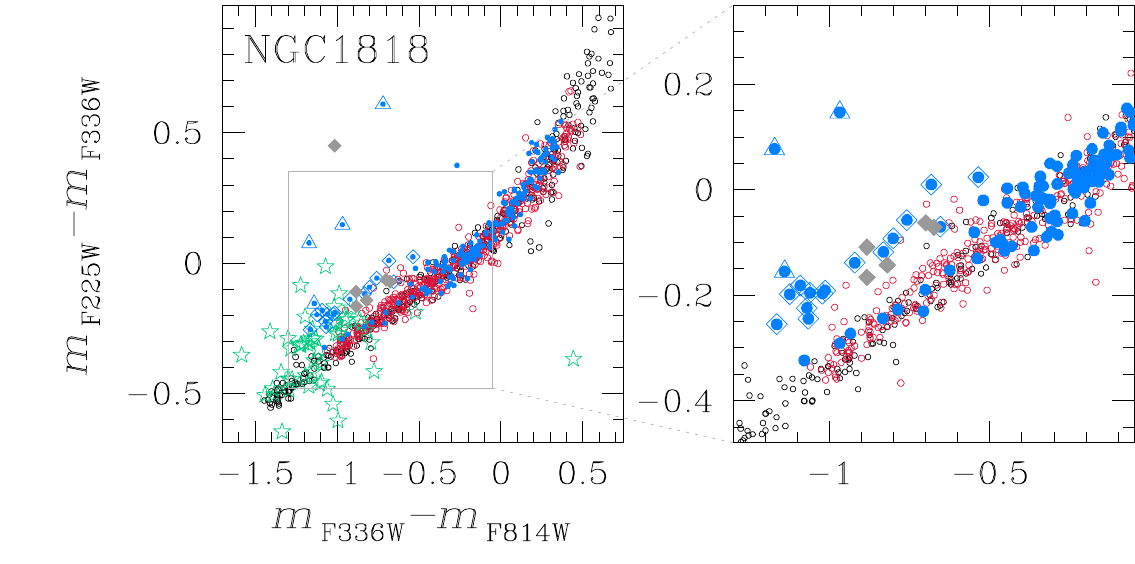}
\includegraphics[height=6.45cm,trim={12.3cm 5.5cm 1.78cm 14.65cm},clip]{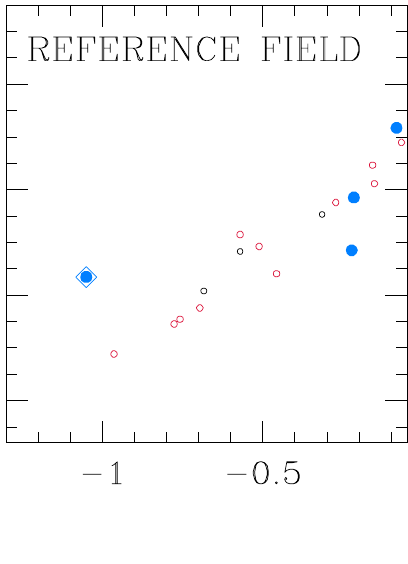}
 \caption{
 $m_{\rm F275W}-m_{\rm F336W}$ ($m_{\rm F225W}-m_{\rm F336W}$)  vs.\,$m_{\rm F336W}-m_{\rm F814W}$ two-color diagrams for stars in the cluster fields of the LMC clusters NGC\,1850, NGC\,1805, and NGC\,1818 (right panels). The middle panels are zooms of the right-panel diagrams for stars in the upper MS. The two-color diagrams plotted in the right panels show the stars in the reference field.
 The symbols and colors are the same as in Figure\,\ref{fig:cmds}. For clearness, in the middle and right panels, we excluded the Be-stars.
 } 
 \label{fig:2Cb} 
\end{figure*}

\begin{figure*} 
\centering
\includegraphics[height=6.45cm,trim={0.0cm 5.5cm 1.78cm 14.65cm},clip]{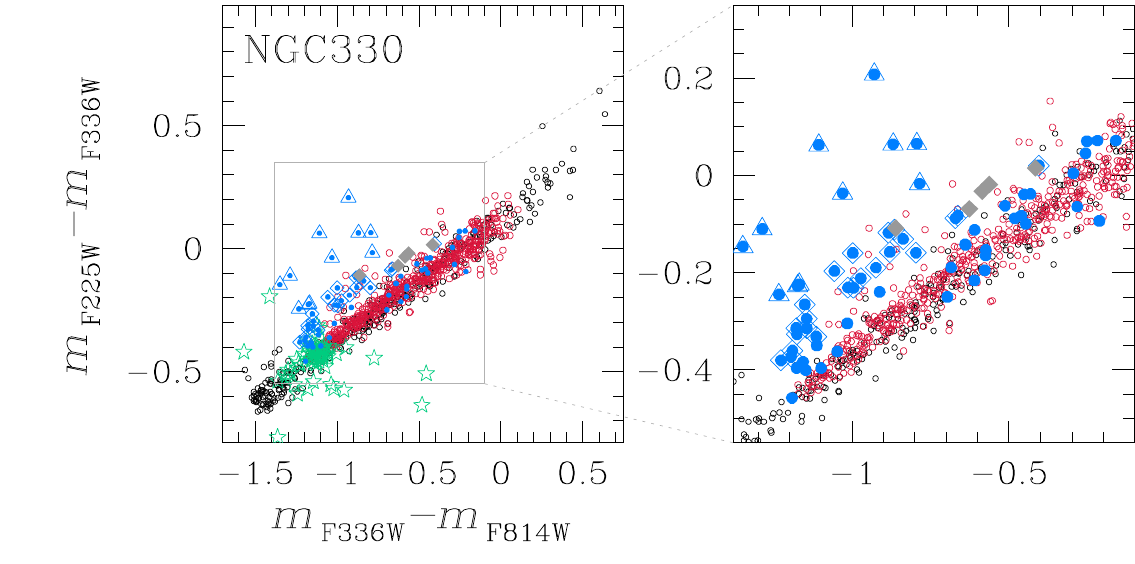}
\includegraphics[height=6.45cm,trim={12.3cm 5.5cm 1.78cm 14.65cm},clip]{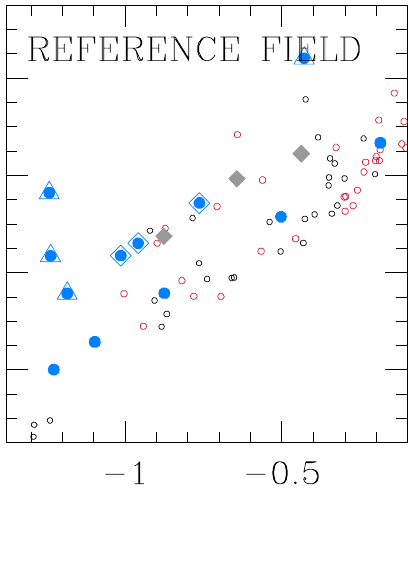}
%
 \caption{As in the Figure\,\ref{fig:2Cb} but for the SMC cluster NGC\,330.}
 \label{fig:2Cc} 
\end{figure*}

To search for UV-dim stars in the other young clusters with available F225W or F275W photometry, we extended the analysis of Section\,\ref{sec:UVn2146} to the LMC clusters NGC\,1850, NGC\,1805, and NGC\,1818, and to the Small Magellanic Cloud (SMC) cluster NGC\,330. The results are illustrated in Figures\,\ref{fig:2Cb} and \ref{fig:2Cc}, where we show the $m_{\rm F225W}-m_{\rm F336W}$ (or $m_{\rm F275W}-m_{\rm F336W}$) vs.\,$m_{\rm F336W}-m_{\rm F814W}$ two-color diagrams for stars in the cluster field (left and middle panels) and in the reference field (right panels).

Similar to what is observed for NGC\,2164, the red-MS stars of these four clusters, which are identified from the $m_{\rm F336W}$ vs.\,$m_{\rm F336W}-m_{\rm F814W}$ CMD, define  narrow sequences in the two-color diagrams of Figures\,\ref{fig:2Cb} and \ref{fig:2Cc}. Conversely, the blue-MS stars with $m_{\rm F336W}-m_{\rm F814W} \lesssim -0.2$ mag are spread over a wide area of the two-color diagram and exhibit larger $m_{\rm F225W}-m_{\rm F336W}$ values than the majority of MS stars with similar $m_{\rm F336W}-m_{\rm F814W}$ colors. 
A visual comparison between the two-color diagrams of stars in the cluster and reference fields shows that contamination from field stars is negligible in all clusters.
We conclude that UV-dim stars are common features in the CMDs of clusters younger than $\sim$200 Myr.

Intriguingly, the Be stars of the different clusters exhibit different behaviors in the two-color diagrams. While nearly all Be stars of NGC\,2164 and NGC\,1850 define narrow sequences in the $m_{\rm F225W}-m_{\rm F336W}$ (or $m_{\rm F275W}-m_{\rm F336W}$) vs.\,$m_{\rm F336W}-m_{\rm F814W}$ two-color diagrams, several Be stars of NGC\,1805, NGC\,330, and NGC\,1818 exhibit a large star-to-star color scatter.

\section{Summary and discussion}\label{sec:summary}
We analyzed UVIS/WFC3 photometry of five Magellanic-Cloud clusters younger than $\sim$200 Myr in the F225W (or F275W), F336W, and F814W bands. We also used F656N photometry to identify Be stars.
We discovered  that a significant fraction of their bright blue-MS stars exhibit larger $m_{\rm F225W}-m_{\rm F336W}$ colors than the bulk of MS stars, in close analogy with what we observed for UV-dim stars in NGC\,1783. 
 The colors of these stars are not consistent with the colors of stellar populations with different rotation rates that we simulated by using the Geneva models \citep{Ekstrom2012, ekstrom2013a}.  
 We conclude that UV-dim stars are not a peculiarity of the $\sim$1.7-old star cluster NGC\,1783 but are common features of young Magellanic-Cloud clusters. 
 Intriguingly, most of the UV-dim stars detected in young clusters populate the bright portion of the blue-MS, which is composed of slow-rotating stars
\citep{Dantona2015a, Marino2018a}.

 In a recent paper, \citet{martocchia2023a} analyzed optical and ultraviolet photometry of NGC\,1850 and suggested that the rapidly-rotating Be stars of this cluster, previously identified by \citet{kamann2023a} by using MUSE spectra, correspond to the UV-dim stars.
They concluded that UV-dim stars are possibly rapidly rotating stars, which are seen nearly equator-on and thus are extinct by their own decretion disks.

Our investigation does not exclude the connection between UV-dim stars and Be stars suggested by these authors.
Nevertheless, since all UV-dim stars that we identified in NGC\,1850 do not exhibit H$\alpha$ emission, they do not correspond to the sample of UV-dim stars by Martocchia and collaborators.  
A significant difference between the bulk of Be stars and the selected UV-dim stars is that they populate distinct regions of the $m_{\rm F225W}-m_{\rm F336W}$ (or $m_{\rm F275W}-m_{\rm F336W}$) vs.\,$m_{\rm F336W}-m_{\rm F814W}$ two-color diagrams.  Specifically, while Be and UV-dim stars share a similar interval of $m_{\rm F336W}-m_{\rm F814W}$, the majority of UV-dim stars are spread over a wide interval of $m_{\rm F225W}-m_{\rm F336W}$.

  The physical mechanisms that are responsible for the UV-dim stars are investigated by \citet{dantona2023a} who compared the observed CMDs of NGC\,1783 stars with simulated photometry by assuming that  some stars are obscured by dust.
Since the amount of absorption due to dust depends on the wavelength, they reproduced the colors of UV-dim stars by assuming that the dust is distributed in a ring and the stars are seen nearly equator-on. 
 D'Antona and collaborators suggested that circumstellar dust is associated with mass-loss events that can preferentially occur in fast-rotating stars.

The fact that the UV-dim stars in young clusters are slow rotators could be a challenge for the scenario by \citet{dantona2023a}. 
However, as proposed by \citet{DAntona2017a}, the blue MSs of young clusters may be composed of stars initially rapidly rotating, but that have later slowed down. 
In this context, we may speculate that the UV-dim stars of the young clusters are indeed stars that have been recently subject to an intense mass loss event. This event caused both their abnormal UV absorption and braked their rotation, shifting the star location to the slow-rotating MS (Dell'Agli et al.\,in preparation). 
Alternatively, if the blue MS is due to merging \citet{wang2022a}, we suggest that the merging event has been recent enough that the strong UV absorption is caused by the remnant circumstellar debris of this event.

\section*{acknowledgments} 
We thank the referee for improving the manuscript.
This work has received funding from the European Union’s Horizon 2020 research and innovation programme under the Marie Sklodowska-Curie Grant Agreement No. 101034319 and from the European Union – NextGenerationEU, beneficiary: Ziliotto.
APM, GC, AFM, FD, FD, and PV acknowledge the support received from
INAF Research GTO-Grant Normal RSN2-1.05.12.05.10 - Understanding the formation of globular clusters with their multiple
stellar generations (ref. Anna F. Marino) of the "Bando INAF per
il Finanziamento della Ricerca Fondamentale 2022"
\small
  
\section*{Data availability}
The data underlying this article will be shared upon reasonable request to the corresponding author.

\bibliography{ms}

\section*{Appendix. Proper motions of the star cluster BHRT\,5b and of LMC stars}

In Section\,\ref{sec:data}, we derived the proper motions of stars in the direction of NGC\,1850 and used them to separate the bulk of cluster members from field stars.
In the following, we use these proper motions to better constrain the kinematics of the stellar populations in the field of view.

\begin{figure*} 
\centering
\includegraphics[height=8.5cm,trim={0cm 0cm 1.cm 0cm},clip]{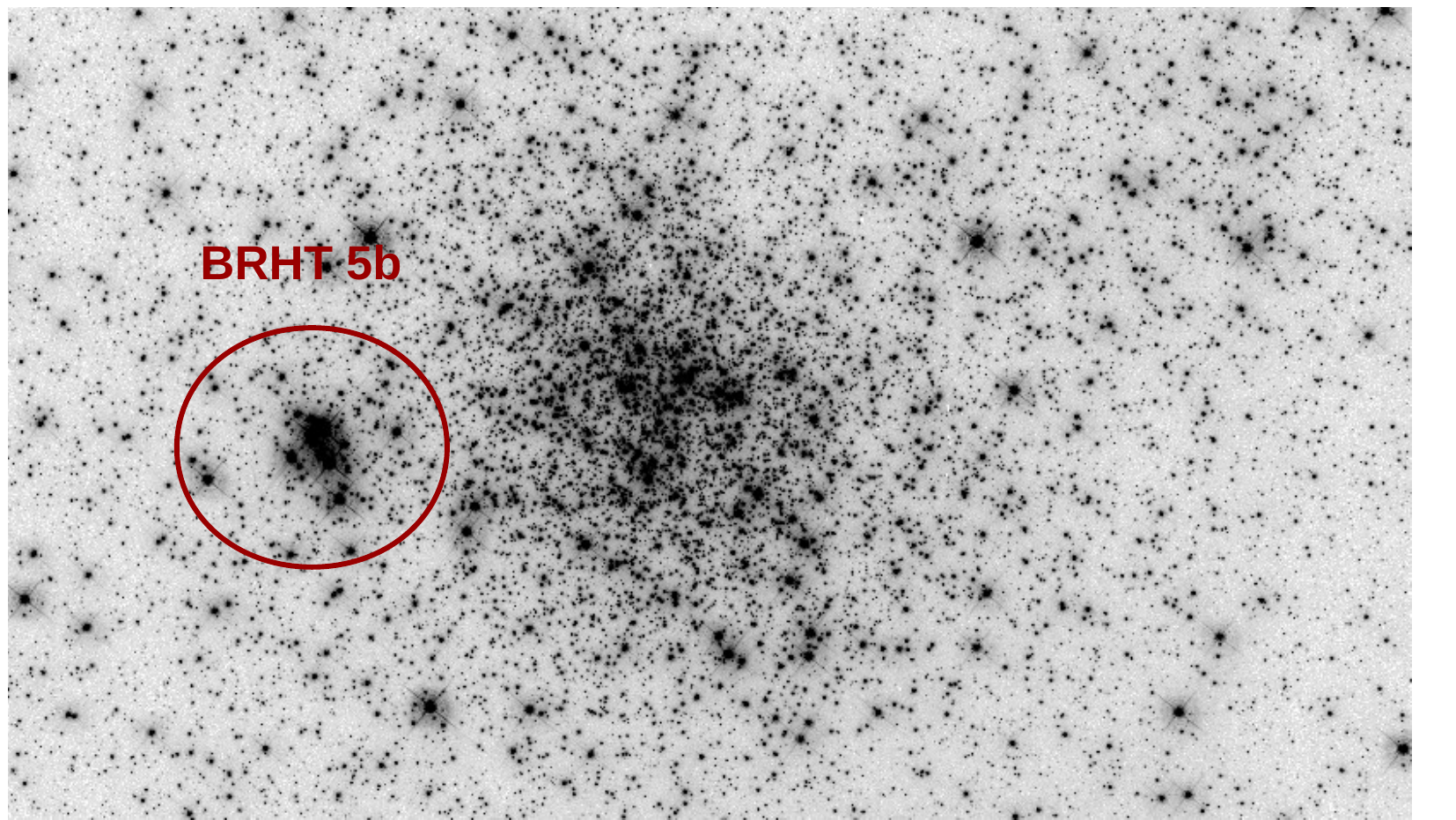}
\includegraphics[height=7.45cm,trim={0.3cm 5.5cm 0cm 14cm},clip]{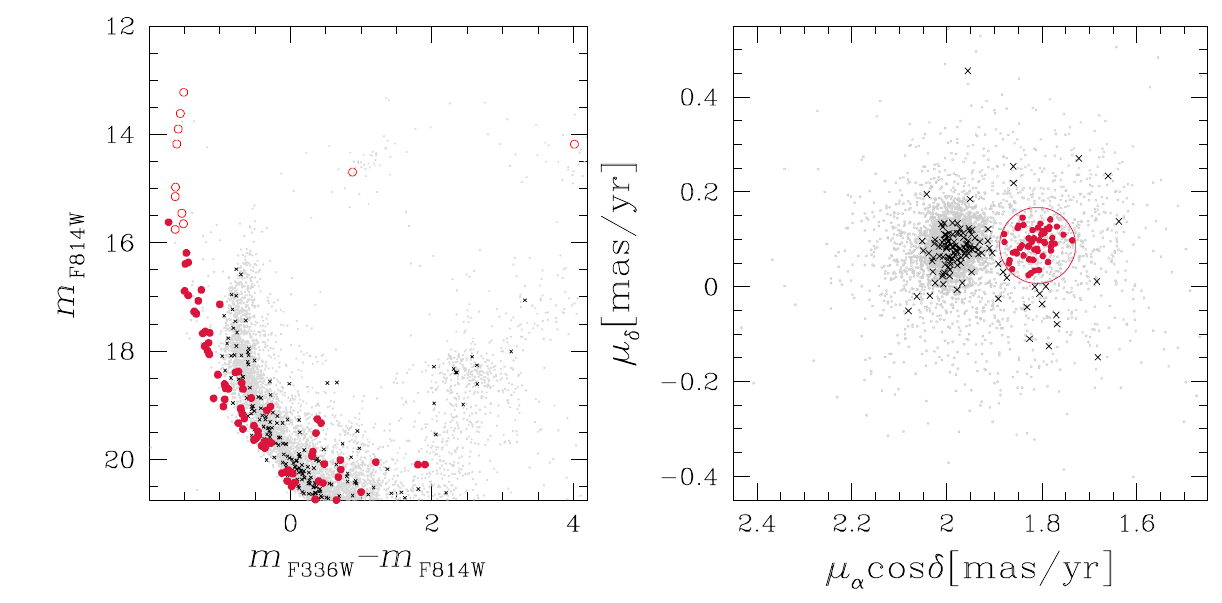}
 \caption{Stacked F467M UVIS/WFC3 image of the star clusters NGC\,1850 and BRHT\,5b (top). The left panels show the $m_{F814W}$ vs.\,$m_{\rm F336W}-m_{\rm F814W}$ CMD (left) and the proper-motion diagram (right) of stars in the field of view of NGC\,1850. The crimson-filled dots show the probable BRHT\,5b members that are located within the crimson circle shown in the top panel and that we selected from the proper motion diagram. The remaining stars with a distance smaller than 12 arcsec from the center of BRHT\,5b, and available proper motions are represented with black crosses. The stars within 12 arcsec from the cluster center and without proper motions are plotted with open circles. }
 \label{fig:BHRT5b} 
\end{figure*}

As shown in the top panel of Figure\,\ref{fig:BHRT5b}, where we plot the stacked  F467M UVIS/WFC3 image of stars around NGC\,1850, a distinctive feature of this cluster is the presence of BHRT\,5b, which is a compact cluster $\sim$30 arcsec west from its center. Due to the small projected distance between these two clusters, NGC\,1850 is considered a double cluster, but the possibility that the two clusters are gravitationally bound is still debated \citep[e.g.][and references therein]{sollima2022a}.

 To investigate the proper motion of BHRT\,5b, we selected the stars within the crimson circle plotted in the bottom panel of Figure\,\ref{fig:BHRT5b}, which is centered on BHRT\,5b and has a radius of 12 arcsec. These stars are plotted with colored symbols in the CMD and proper motion diagrams of the lower panels.
 
  As shown in the bottom-right panel of Figure\,\ref{fig:BHRT5b}, most of the selected stars are clustered into two regions of the proper motion diagram. Some stars share the same proper motions of NGC\,1850 and have similar colors and magnitudes thus indicating that they are probable members of this cluster. Conversely, the group of stars that we associate to BHRT\,5b is clustered around ($\mu_{\alpha} cos \delta$, $\mu_{\delta}$)=(1.822$\pm$0.008$\pm$0.026,0.087$\pm$0.007$\pm$0.028) mas/yr. Here, the two error components indicate the uncertainty on the proper motion relative to NGC\,1850 and the error on the absolute proper motion of NGC\,1850. 
  To select the probable members of BHRT\,5b in the proper motion diagram we iteratively defined a circle centered on the BHRT\,5b average motion and with a radius of 0.08 mas/yr, which corresponds to 3 times the average proper motion uncertainty.
  The fact that these stars, which are colored with crimson-filled dots, populate the brightest MS portion of the CMD corroborates the evidence that they belong to BHRT\,5b. We conclude that, due to the different proper motions, NGC\,1850 and BHRT\,5b are not gravitationally bound. 

\begin{figure*} 
\centering
\includegraphics[height=7.0cm,trim={0cm 0cm 1.cm 0cm},clip]{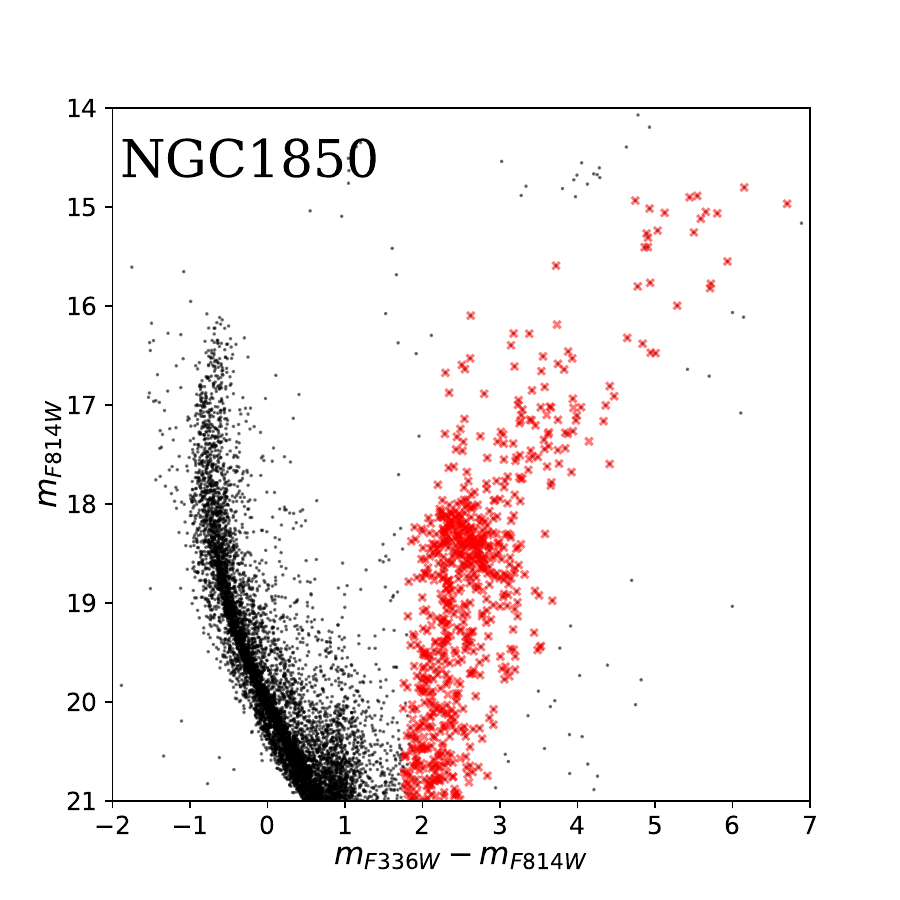}
\includegraphics[height=7.0cm,trim={0cm 0cm 1cm 0cm},clip]{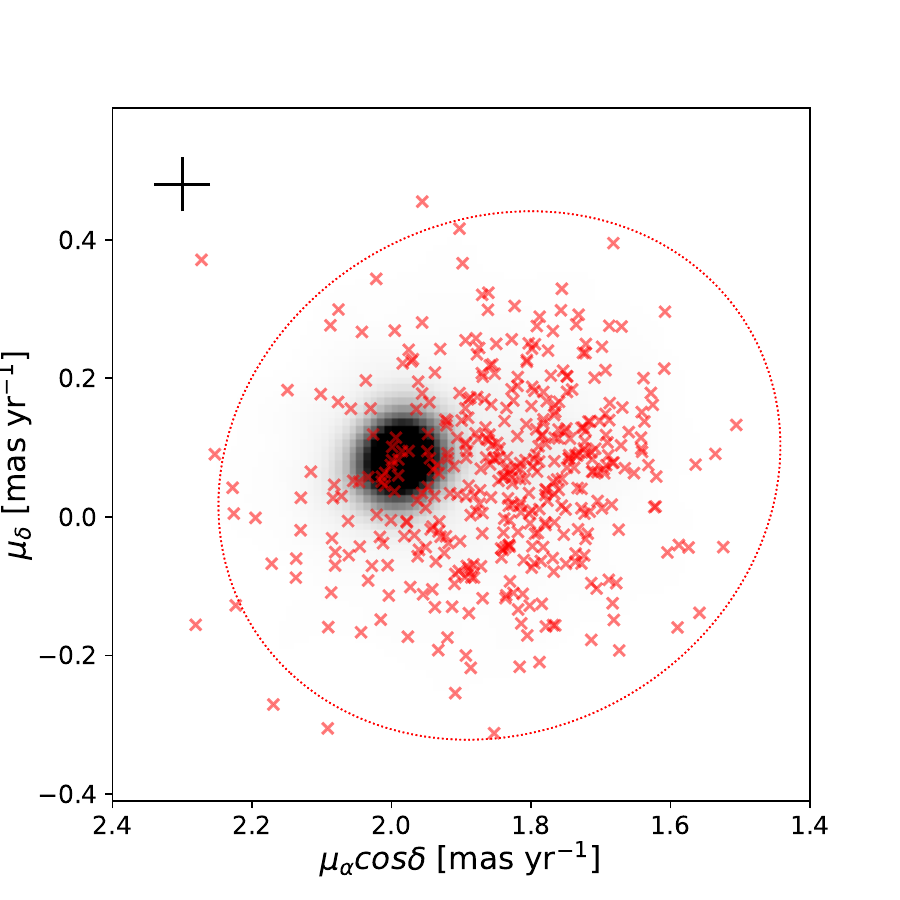}
 \caption{$m_{F814W}$ vs.\,$m_{\rm F336W}-m_{\rm F814W}$ CMD (left) and proper-motion diagram of stars in the field of view of NGC\,1850. The LMC RGB stars are marked with red crosses, while the red ellipse provides the best fitting of the proper-motion distribution of RGB stars.}
 \label{fig:ngc1850pm2} 
\end{figure*}

  In addition to constraining the kinematics of star clusters, the stellar proper motions allow us to characterize the stellar populations of the LMC. 
  To do that, we used the CMD plotted in left panel of Figure\,\ref{fig:ngc1850pm2}  to identify the LMC RGB stars in the field of view of NGC\,1850 (red crosses). As shown in the right panel of Figure\,\ref{fig:ngc1850pm2}, the selected LMC stars have a median proper motion ($\mu_{\alpha} cos \delta$, $\mu_{\delta}$)=(1.831$\pm$0.006$\pm$0.026,0.065$\pm$0.006$\pm$0.028) mas/yr and
 exhibit a nearly circular distribution. Specifically, the ellipse that provides the best fit with the observed proper motion distribution has ellipticity, $\epsilon=0.10 \pm 0.07$ and position angle, $\theta=37 \pm 12$ degrees \footnote{We considered the ellipse that encloses 90\% of stars, and defined the ellipticity $\epsilon=1-b/a$, where $a$ and $b$ are the minor and major axes of the ellipse \citep[see][for details]{milone2023a}.}.

  The evidence that the old LMC stars in the background and foreground NGC\,1850 exhibit nearly circular distributions is consistent with the findings by \citet{milone2023a} based on LMC stars in the field of views of NGC\,1755, NGC\,1801, and NGC\,1953.  In contrast, the behaviour of the proper motions of old LMC stars is different from what observed for both young and old SMC stars, whose proper motions exhibit high ellipticity values and major axes that point toward the LMC \citep{milone2023a, zivick2018a, dias2021a, piatti2021a, schmidt2022a}. 

\end{document}